\documentclass[12pt]{article}

\catcode`\@=11
\@addtoreset{equation}{section}

\global\arraycolsep=2pt
\usepackage{amsbsy,amssymb,latexsym,amsfonts,amsmath}
\usepackage{graphicx,color}

\allowdisplaybreaks

\begin{document}
\begin{flushright}
\parbox{4.2cm}
{UCB-PTH-09/07}
\end{flushright}

\vspace*{0.7cm}

\begin{center}
{\Large \bf 
Superfield Formulation 

for Non-Relativistic Chern-Simons-Matter Theory
}
\vspace*{2.0cm}\\
{Yu Nakayama}
\end{center}
\vspace*{-0.2cm}
\begin{center}
{\it Berkeley Center for Theoretical Physics, \\ 
University of California, Berkeley, CA 94720, USA
}
\vspace{3.8cm}
\end{center}

\begin{abstract} 
We construct a superfield formulation for non-relativistic Chern-Simons-Matter theories with manifest dynamical supersymmetry. By eliminating all the auxiliary fields, we show that the simple action reduces to the one obtained by taking non-relativistic limit from the relativistic Chern-Simons-Matter theory proposed in the literature. As a further application, we give a manifestly supersymmetric derivation of the non-relativistic ABJM theory. 
\end{abstract}

\thispagestyle{empty} 

\setcounter{page}{0}

\newpage

\section{Introduction} 
The recent advent of the non-relativistic AdS/CFT correspondence \cite{Son:2008ye}\cite{Balasubramanian:2008dm} makes it more important than recognized before to study non-relativistic superconformal gauge theories. It may have realistic applications to strongly coupled condensed matter physics such as high $T_c$ superconductor or quantum Hall effects. On the other hand, theoretical understanding of the AdS/CFT correspondence may be enhanced by the comparison of such theoretical predictions with experiments.

In particular, non-relativistic Chern-Simons-Matter theories in $(1+2)$ dimension are of special relevance. The Galilean invariance as well as non-relativistic conformal invariance (known as Schr\"odinger invariance \cite{Hagen:1972pd}\cite{Niederer:1972zz}\cite{Perroud:1977qh}\cite{Barut:1981mt}) can be implemented by directly taking the non-relativistic limit of relativistic Chern-Simons-Matter theory in the same dimension \cite{Jackiw:1990mb}\cite{Jackiw:1992fg}. The construction can be applied to the supersymmetric Chern-Simons-Matter theory as well \cite{Leblanc:1992wu}, and explicit examples of non-relativistic superconformal field theories have been constructed in this way \cite{Nakayama:2008qz}\cite{Nakayama:2008td}. As a particular example, various non-relativistic limits of the M2-brane gauge theory  were scrutinized in \cite{ABJM},\footnote{Our construction is to deform the original ABJM theory \cite{Aharony:2008ug}, which corresponds to the M2-brane gauge theory on the orbifold space, by adding the supersymmetric mass term \cite{Hosomichi:2008jb}\cite{Gomis:2008vc}, which corresponds to the introduction of four-form flux, and take the subsequent non-relativistic limit.} which may admit the explicit gravity solution that can be studied in the context of non-relativistic AdS/CFT correspondence.

However, in this approach, the supersymmetry is not manifest, and there is no general argument how to guarantee the supersymmetry preserved in  the course of taking the non-relativistic limit. Indeed, we have seen several counterintuitive examples that break the supersymmetry \cite{Nakayama:2008qz}\cite{Nakayama:2008td} through this blind, seemingly harmless, non-relativistic limit. Thus, we have to check the supersymmetry of the proposed non-relativistic action by hand, and sometimes we have to add some other terms, like four-fermi terms, to restore the supersymmetry \cite{Nakayama:2008td}.  In particular, it is very cumbersome to verify the dynamical supersymmetry in the component formulation.

Superfield formulation \cite{Salam:1974yz} (see e.g. \cite{Gates:1983nr}\cite{Wess:1992cp} for reviews in the relativistic case) is an elegant way to make the supersymmetry transformation manifest. With the superfield formulation at hand, the supersymmetry transformation is nothing but the translation in the superspace, and the construction of the supersymmetric action becomes simply the total integral over the superspace. In this paper, we develop a superfield formulation for the non-relativistic Chern-Simons-Matter theory (see also \cite{deAzcarraga:1991fa} for another superfield formulation for non-relativistic systems), and we propose a manifest superfield form of the action studied in the literatures.

Our construction is minimal in the sense that we only assume the dynamical supersymmetry realized by the algebra
\begin{align}
\{Q_2^*, Q_2\} = 2H \ ,
\end{align}
so we can use our formulation not only to particular Chern-Simons-Matter theories but also any other supersymmetric theories involving this algebra: even conventional relativistic supersymmetric field theories could be rewritten in our formulation while some of the Poincar\'e symmetry would not  be manifest. We may also use our formulation to construct a supersymmetric extension of the Lifshitz-Ho\v{r}ava non-relativistic gauge theory \cite{Horava:2008jf}.

The organization of the paper is as follows. In section 2, we introduce the superfield formulation of the non-relativistic gauge theory in $(1+2)$ dimension. In section 3, we re-examine the non-relativistic Abelian ABJM theory in our formulation.  In section 4, we generalize our construction to the non-Abelian case, and re-derived the action for the non-relativistic ABJM theory  with 14 supercharges. In section 5, we conclude the paper with further discussions.

\section{Superfield Formulation}
In this section, we develop a superfield formulation of the non-relativistic Abelian gauge theory coupled with matters. Our goal is to provide a manifest supersymmetric action for the non-relativistic Chern-Simons-Matter theory introduced in \cite{Leblanc:1992wu}.

\subsection{Superspace and superfields}
In order to realize the dynamical supersymmetry 
\begin{eqnarray}
\{Q_2^*, Q_2\} = 2H \label{alge}
\end{eqnarray}
in a manifest form, we introduce the superspace $(t,x^i, \theta , \bar{\theta})$, where $\theta$ and $\bar{\theta}$ are one-component complex Grassmann coordinate.\footnote{We always work in $1+2$ dimension, so $i=1,2$.} We can realize the superalgebra \eqref{alge} by the supercharge
\begin{eqnarray}
\mathcal{Q} = \frac{\partial}{\partial \theta} + i \bar{\theta}\partial_t \ ,  \ \ \bar{\mathcal{Q}} = \frac{\partial}{\partial \bar{\theta}} + i \theta \partial_t \ ,
\end{eqnarray}
which satisfy the anti-commutation relation $\{\mathcal{Q}, \bar{\mathcal{Q}}\} = 2i\partial_t$. We also introduce the supersymmetric derivative:
\begin{eqnarray}
 D =  \frac{\partial}{\partial \theta} - i \bar{\theta}\partial_t \ , \bar{D} = \frac{\partial}{\partial \bar{\theta}} - i \theta \partial_t \ ,
\end{eqnarray}
which satisfy the anti-commutation relation  $\{D, \bar{D}\} = -2i\partial_t$.

The most general superfield $\Sigma(t,x^i,\theta,\bar{\theta})$ that lives on the superspace can be expanded as
\begin{eqnarray}
\Sigma = a + \theta b + \bar{\theta} c + \theta \bar{\theta} d \ ,
\end{eqnarray}
where $a$ and $d$ are complex bosons and $b$ and $c$ are complex one-component fermions. By acting $\epsilon \mathcal{Q}$ and $\epsilon^* \bar{\mathcal{Q}}$, the supersymmetry transformation can be read as
\begin{align}
\delta_\epsilon a &= \epsilon b  \ , & \delta_{\epsilon^*}a &= \epsilon^* c \cr
\delta_\epsilon b &= 0    \ , & \delta_{\epsilon^*} b&= \epsilon^* (d-i\partial_t a) \cr
\delta_\epsilon c &= -\epsilon (d + i\partial_t a)  \ , & \delta_{\epsilon^*}c &=0  \cr
\delta_\epsilon d &= -i\epsilon \partial_t b \ , & \delta_{\epsilon^*}d &= i\epsilon^* \partial_t c \ .  
\end{align}
where $\epsilon, \epsilon^*$ are Grassmann supersymmetry parameters. 
We note that the $\theta \bar{\theta}$ component $d$ transforms as a total derivative, which will be crucial to construct supersymmetric actions.

In order to reduce the degrees of freedom contained in a general superfield, constrained superfields will be useful for our studies.\footnote{For a time being, we assume that the top component of the superfields is bosonic. We will later introduce fermionic superfields as well.} One simple choice is to impose the reality condition $ \Upsilon = \Upsilon^*$ (real superfield): in component, we have
\begin{eqnarray}
\Upsilon = a + \theta b - \bar{\theta} {b}^* + \theta \bar{\theta} d \ ,
\end{eqnarray}
where $a$ and $d$ are now real bosons and $b$ is a complex fermion.

Another simple constrained superfield is a chiral superfield $X$ satisfying $\bar{D} X = 0$: in component, we have
\begin{eqnarray}
X = x + \theta \chi - \theta \bar{\theta} (i \partial_t x) \ , 
\end{eqnarray}
where $x$ is a complex boson and $\chi$ is a complex fermion.

Similarly anti-chiral superfield is obtained by requiring the condition $D\bar{X} = 0$, whose component form is
\begin{eqnarray}
\bar{X} = x^* - \bar{\theta} \chi^* + \theta \bar{\theta} i \partial_t x^* \ .
\end{eqnarray}
Note that $X$ and $\bar{X}$ are complex conjugate with each other: $X^* = \bar{X}$.

\subsection{Gauge multiplet}
To construct supersymmetric non-relativistic Chern-Simons-Matter theories, we begin with a gauge multiplet. It turns out that the gauge multiplet consists of one real superfield $\mathcal{V}$ and one chiral superfield $\mathcal{A}$ (and its complex conjugate ${\mathcal{A}}^*$). We will see that $\mathcal{V}$ includes the time component of the gauge field $A_0$ and $\mathcal{A}$ includes the space components $A_i$.

A real superfield $\mathcal{V} = a + \theta \kappa -\bar{\theta} \kappa^* + \theta\bar{\theta} 2A_0 $ has a supersymmetric gauge transformation 
\begin{eqnarray}
\mathcal{V} \to \mathcal{V} + \Lambda + \bar{\Lambda} \ 
\end{eqnarray}
with a supersymmetric gauge parameter given by a chiral superfield $\Lambda = \tilde{\Lambda} + \theta \tilde{\lambda} - \theta \bar{\theta} (i\partial_t \tilde{\Lambda})$. In component, the supersymmetric gauge transformation takes the form
\begin{align}
a &\to a + \tilde{\Lambda} + \tilde{\Lambda}^* \cr
\kappa &\to \kappa + \tilde{\lambda} \cr
2A_0 &\to 2A_0 - i\partial_t\tilde{\Lambda} + i \partial_t\tilde{\Lambda}^* \ .
\end{align}
It is very convenient to use the WZ-gauge, where $a = \kappa = 0$. In this gauge, the supersymmetric gauge transformation is nothing but the ordinary gauge transformation for the time component of the gauge field: $A_0 \to A_0 + \partial_t \mathrm{Im} \tilde{\Lambda}$. 

The chiral superfield $\mathcal{A} = 2A + \theta \lambda - \theta \bar{\theta} (2i\partial_t A)$ includes a complex combination of the gauge field $A = A_1 + iA_2$. We impose the supersymmetric gauge transformation
\begin{eqnarray}
\mathcal{A} \to \mathcal{A}- 2i\partial_+ \Lambda \ ,
\end{eqnarray}
where $\partial_+ = \partial_1 + i\partial_2$ (and similarly we define $\partial_- = \partial_1 - i\partial_2$). It reduces to the usual gauge transformation $2A \to 2A + 2\partial_+(\mathrm{Im} \tilde{\Lambda})$ in the WZ-gauge. Similarly, the anti-chiral superfield $\mathcal{A}^* = 2A^* - \bar{\theta} \lambda^* + \theta \bar{\theta}(2i\partial_t A^*)$ has the supersymmetric gauge transformation $\mathcal{A}^* \to \mathcal{A}^* + 2i\partial_- \bar{\Lambda}$. Note that the ``gaugino" $\lambda$ cannot be eliminated in the WZ-gauge, but we will see that it is an auxiliary field in the Chern-Simons-Matter action.

We propose the supersymmetric extension of the Chern-Simons action as
\begin{eqnarray}
S_{\mathrm{CS}} = -\frac{\kappa}{16} \int dt d^2x d^2\theta \left((\mathcal{A}+2i\partial_+\mathcal{V})(\mathcal{A}^* -2i\partial_- \mathcal{V}) + 2\mathcal{V} \partial_+ \partial_- \mathcal{V} \right) \ , \label{csaction}
\end{eqnarray}
where the convention of the superspace integration is $\int d^2\theta (\theta \bar{\theta}) = 1$. Because of the superspace integration, the supersymmetric variation of the Lagrangian density is a total derivative and the action is invariant under the dynamical supersymmetry. In addition, the supersymmetric gauge transformation of the Lagrangian density is a total derivative as usual in the Chern-Simons theory, so the action is gauge invariant.

In the WZ-gauge, we can evaluate the action \eqref{csaction} as
\begin{eqnarray}
 \int dt d^2x \left(\kappa A_0F_{12} + \frac{\kappa}{2}  \epsilon^{ij}\partial_tA_i A_j -\frac{\kappa}{16} \lambda \lambda^*  \right) \ .
\end{eqnarray}
This is the conventional form of the Abelian Chern-Simons action in the non-relativistic form.

\subsection{Matter multiplet}
Now, we would like to couple the matter multiplets to the Chern-Simons action. For this purpose, we introduce a chiral superfield 
\begin{eqnarray}
\Phi = \phi + \theta \chi - \theta \bar{\theta}(i\partial_t \phi) \ .
\end{eqnarray}
The chiral superfield has the supersymmetric gauge transformation 
\begin{eqnarray}
\Phi \to e^{-\Lambda} \Phi \ , 
\end{eqnarray}
or similarly, the conjugate anti-chiral superfield $\Phi^*$ has the supersymmetric gauge transformation
\begin{eqnarray}
\Phi^* \to e^{-\bar{\Lambda}} \Phi^* \ .
\end{eqnarray}
 
A simple choice of the action
\begin{eqnarray}
S_0 = -\frac{1}{2} \int dt d^2x d^2 \theta \Phi^* e^{\mathcal{V}}\Phi
\end{eqnarray}
is gauge invariant and supersymmetric, but it does not contain any kinetic term for $\phi$:\footnote{This is because our formulation lacks the manifest Galilean invariance.}
\begin{eqnarray}
S_0 = \int dt d^2 x \left( i\phi^*(\partial_0 + iA_0)\phi + \frac{1}{2}\chi^* \chi \right)\ ,
\end{eqnarray}
 so we have to introduce other ingredients. For this purpose, we introduce an associated fermionic chiral superfield $\Psi$, and its conjugate $\Psi^{*}$:
\begin{align}
\Psi &= \psi + \theta \rho - \theta \bar{\theta}(i\partial_t \psi) \cr
\Psi^*&= \psi^* + \bar{\theta} \rho^* + \theta \bar{\theta}(i\partial_t \psi^*) \ ,
\end{align}
where $\psi$ is a complex fermion and $\rho$ is a complex scalar. The gauge transformation is $\Psi \to e^{\Lambda} \Psi$ and $\Psi^* \to e^{\bar{\Lambda}} \Psi^*$.

Now, we impose the supersymmetric covariant constraint between $\Phi$ and $\Psi$:
\begin{eqnarray}
D(e^{\mathcal{V}} \Phi) = \left(\partial_- + \frac{i\mathcal{A}^*}{2}\right) \Psi^* \ .
\end{eqnarray}
The constraint is consistent with the supersymmetry and the gauge symmetry. In components (in the WZ gauge), we have the following constraint
\begin{align}
\chi &= (\partial_- + iA^*)\psi^* \cr
-2(i\partial_t - A_0)\phi &= (\partial_- + iA^*)\rho^* -\frac{i}{2}\lambda^* \psi^* \ . \label{constra}
\end{align}

With this constraint, we proposed a supersymmetric action for the non-relativistic Chern-Simons-Matter theory:
\begin{eqnarray}
S = S_{\mathrm{CS}} -\frac{1}{2}\int dt dx^2 d^2\theta \left(\Phi^* e^{\mathcal{V}} \Phi + \Psi^* e^{-\mathcal{V}} \Psi \right) \ ,
\end{eqnarray}
where $S_{\mathrm{CS}}$ is given in \eqref{csaction}.
In the WZ gauge, the component form of the action can be evaluated as
\begin{align}
S = \int dt dx^2 &\left(\kappa A_0F_{12} + \frac{\kappa}{2} \epsilon^{ij}\partial_t A_i A_j -\frac{\kappa}{16} \lambda \lambda^*  \right. \cr
 &+  i\phi^*(\partial_0 + iA_0)\phi + \frac{1}{2}\chi^* \chi + i\psi^*(\partial_0 - iA_0)\psi + \frac{1}{2}\rho^*\rho \cr
&+ b(\chi - (\partial_- + iA^*) \psi^*) - {b}^*(\chi^* - (\partial_+ - iA) \psi)  \cr
&+ c\left[(i\partial_t - A_0)\phi + \frac{1}{2}(\partial_- + iA^*)\rho^* -\frac{i}{4}\lambda^* \psi^* \right] \cr
& \left. +  c^*\left[(-i\partial_t - A_0)\phi^* + \frac{1}{2}(\partial_+ - iA)\rho +\frac{i}{4}\psi\lambda \right] \right) \ . \label{full}
\end{align}
Here, in the last three lines, we have introduced a fermionic Lagrange multiplier $b$ and a bosonic Lagrange multiplier $c$ to impose the constraint \eqref{constra}.

The action is invariant under the dynamical supersymmetry. It is less obvious but it is also invariant under the kinematical supersymmetry
\begin{eqnarray}
\{ Q_1 , Q_1^*\} = 2M \ , \ \ \{Q_1, Q_2^*\} = P_- \ ,
\end{eqnarray}
where $M$ is the total mass operator and $P_-$ is the momentum operator.
Alternatively speaking, the action has a Galilean invariance, so the commutator $i[{G}_-,Q_2] = -Q_1$ guarantees the existence of the kinematical supersymmetry (see \cite{Leblanc:1992wu} for $\mathcal{N}=2$ non-relativistic supersymmetry algebra).
Indeed, we will show that the action \eqref{full} is equivalent to the one proposed in \cite{Leblanc:1992wu} as a non-relativistic limit of $\mathcal{N}=2$ Chern-Simons-Matter theory.

\subsection{Equivalence to \cite{Leblanc:1992wu}}
In order to show the equivalence to the action proposed in \cite{Leblanc:1992wu}, we first use the equation of motions for $\lambda$, $\rho$ and $\phi$ 
\begin{align}
\frac{\kappa}{16} \lambda - i\frac{c}{4} \psi^* &= 0 \cr
\frac{\rho}{2} + \frac{1}{2}(-\partial_- + iA^*) c &= 0 \cr
i(\partial_t + iA_0) \phi + i(\partial_t+iA_0)c^* & = 0 \label{eom} 
\end{align}
to eliminate $\lambda$, $\rho$ and $\phi$ (as well as $\chi$ from the constraint). Although the last equation in \eqref{eom} is not kinematical but dynamical, it is easy to see that $\phi = -c^*$ is the solution.

Substituting them into the action \eqref{full}, we have 
\begin{align}
S = \int dt dx^2 &\left(\kappa A_0F_{12} + \frac{\kappa}{2} \epsilon^{ij}\partial_t A_i A_j  \right. \cr
 &+  ic^*(\partial_0 - iA_0)c - \frac{1}{2}(\partial_+ + iA)c^*(\partial_- - iA^*)c  \cr
&+ i\psi^*(\partial_0 - iA_0)\psi -\frac{1}{2}(\partial_- + iA^*)\psi^* (\partial_+-iA)\psi +\frac{c^*c\psi^*\psi}{\kappa}  \label{eqaction}
\end{align}
In order to compare the action with that in \cite{Leblanc:1992wu} we further use the Gauss-law constraint
\begin{eqnarray}
F_{12} = -\frac{1}{\kappa} (c^*c + \psi^* \psi) \ ,
\end{eqnarray}
and rewrite the kinetic terms by using a  trick \cite{Nakayama:2008qz} (up to total derivative terms)
\begin{align}
-\frac{1}{2}(\partial_+ + iA)c^*(\partial_- - iA^*)c  &= -\frac{1}{2}(\partial_i+iA_i)c^*(\partial_i-iA_i) -\frac{F_{12}}{2}c^*c \cr
& = -\frac{1}{2}(\partial_i+iA_i)c^*(\partial_i -iA_i)c + \frac{(c^*c)^2 + c^*c\psi^*\psi}{2\kappa} \cr
-\frac{1}{2}(\partial_- + iA^*)\psi^* (\partial_+-iA)\psi &= -\frac{1}{2}(\partial_i+iA_i)\psi^*(\partial_i -iA_i)\psi +\frac{F_{12}}{2}\psi^*\psi
\end{align}

Then, the action can be transformed into
\begin{align}
S = \int dt dx^2 &\left(\kappa A_0F_{12} + \frac{\kappa}{2} \partial_t \epsilon^{ij}A_i A_j  \right. \cr
 &+  i\Phi^*(\partial_0 - iA_0)\Phi - \frac{1}{2}(\partial_i +iA_i)\Phi^*(\partial_i-iA_i)\Phi  \cr
&+ i\Psi^*(\partial_0 - iA_0)\Psi -\frac{1}{2}(\partial_i+ iA_i)\Psi^* (\partial_i-iA_i)\Psi + \frac{F_{12}}{2}\Psi^*\Psi\cr
&\left. +\frac{(\Phi^*\Phi)^2}{2\kappa} + 3\frac{\Phi^*\Phi\Psi^*\Psi}{2\kappa} \right)  
\end{align}
by renaming $c \to \Phi$ and $\psi \to \Psi$.
This action is equivalent to the one presented in \cite{Leblanc:1992wu} with $e=1$ and $m=1$.\footnote{We can easily recover the mass parameter $m$ by considering the constraint $D(e^\mathcal{V}\Phi) = \frac{1}{\sqrt{m}}\left(\partial_- + \frac{i\mathcal{A}^*}{2}\right) \Psi^*$.}

It is invariant under the dynamical supersymmetry:
\begin{align}
\delta \Phi &= \frac{i}{\sqrt{2}} \epsilon^* D_+ \Psi \cr
\delta \Psi &= -\frac{i}{\sqrt{2}} \epsilon D_- \Phi \cr
\delta A &= \frac{2}{\sqrt{2}\kappa} \epsilon \Psi^*\Phi \cr
\delta {A}^* &= -\frac{2}{\sqrt{2}\kappa} \epsilon^* \Psi\Phi^* \cr
\delta A_0 &= \frac{i}{2\sqrt{2} \kappa}[\epsilon(D_- \Psi^*) \Phi +\epsilon^*
(D_+\Psi)\Phi  ]
\end{align}
as well as the kinematical supersymmetry
\begin{align}
\delta \Phi &= \sqrt{2} \eta^* \Psi \cr
\delta \Psi &= -\sqrt{2} \eta \Phi \cr
\delta A_i &= 0 \cr
\delta A_0 & = \frac{1}{\sqrt{2}\kappa}(\eta \Psi^*\Phi
-\eta^*\Psi\Phi^*) \ .
\end{align}

\section{Non-relativistic Abelian ABJM theory}
As a simple application of our superfield formulation, we would like to construct the non-relativistic Abelian ($U(1)\times U(1)$) ABJM theory without referring to the relativistic action. We introduce two vector multiplets ($\mathcal{V}, \mathcal{A}, \mathcal{A}^*$) and ($\hat{\mathcal{V}}, \hat{\mathcal{A}}, \hat{\mathcal{A}}^*$) together with $4$ matter chiral multiplets $(\Phi_A, \Psi_A)$ ($A=1,\cdots,4$). The supersymmetric gauge transformations are
\begin{align}
\mathcal{V} &\to \mathcal{V} + \Lambda + \bar{\Lambda} \cr
\mathcal{A} &\to \mathcal{A} -2i \partial_+ \Lambda \cr
\hat{\mathcal{V}} &\to \hat{\mathcal{V}} + \hat{\Lambda}+\bar{\hat{\Lambda}} \cr
\hat{\mathcal{A}} &\to \hat{\mathcal{A}} -2i\partial_+ \hat{\Lambda} \cr
\Phi_A & \to e^{-\Lambda + \hat{\Lambda}} \Phi_A \cr
\Psi_A & \to e^{\Lambda - \hat{\Lambda}} \Psi_A \  .
\end{align}

The invariant action that we propose for the Abelian non-relativistic ABJM theory is 
\begin{align}
S = \int dt d^2x d^2\theta &\left[-\frac{\kappa}{16} \left((\mathcal{A}+2i\partial_+\mathcal{V})(\mathcal{A}^* -2i\partial_-\mathcal{V}) + 2\mathcal{V} \partial_+ \partial_- \mathcal{V} \right) \right.\cr
 & + \frac{\kappa}{16} \left((\hat{\mathcal{A}}+2i\partial_+\hat{\mathcal{V}})(\hat{\mathcal{A}}^* -2i\partial_- \hat{\mathcal{V}}) + 2\hat{\mathcal{V}} \partial_+ \partial_- \hat{\mathcal{V}} \right) \cr
 & \left. -\frac{1}{2} \Phi_A^* e^{\mathcal{V}-\hat{\mathcal{V}}} \Phi_A + \Psi^*_Ae^{-\mathcal{V}+\hat{\mathcal{V}}} \Psi_A \right]
\end{align}
with the supersymmetric constraint 
\begin{eqnarray}
D(e^{\mathcal{V}-\hat{\mathcal{V}}} \Phi_A) = \left(\partial_- +\frac{i}{2}(\mathcal{A}^* - \hat{\mathcal{A}}^*)\right) \Psi^*_A \ .
\end{eqnarray}

Just exactly as we did in section 2, we can eliminate all the auxiliary fields in the components:
\begin{align}
\frac{\kappa}{16} \lambda - i\frac{c_A}{4} \psi^*_A &= 0 \cr
\frac{\kappa}{16} \hat{\lambda} - i\frac{c_A}{4} \psi^*_A &= 0 \cr
\frac{\rho_A}{2} + \frac{1}{2}(-\partial_- + iA^* -i\hat{A}^*) c_A &= 0 \cr
i(\partial_t + iA_0-i\hat{A}_0) \phi_A + i(\partial_t+iA_0-i\hat{A}_0)c^*_A & = 0 
\end{align}
where $c_i$ are Lagrange multipliers as before.
Substituting them back into the action, we obtain
\begin{align}
S = \int dt dx^2 &\left(\kappa A_0F_{12} + \frac{\kappa}{2} \epsilon^{ij}\partial_tA_i A_j  -\kappa\hat{A}_0\hat{F}_{12} - \frac{\kappa}{2} \epsilon^{ij}\partial_t \hat{A}_i\hat{A}_j \right. \cr
 &+  ic^*_A(\partial_0 - iA_0 + i\hat{A}_0)c_A - \frac{1}{2}(\partial_+ + iA -i\hat{A})c^*_A(\partial_- - iA^* + i\hat{A})c_A  \cr
&\left. + i\psi^*_A(\partial_0 - iA_0 + i\hat{A}_0)\psi_A -\frac{1}{2}(\partial_- + iA^*-i\hat{A}^*)\psi^*_A (\partial_+-iA+i\hat{A})\psi_A \right) 
\end{align}

We can further rewrite the action by using the Gauss-law constraint:
\begin{eqnarray}
F_{12} = \hat{F}_{12} = -\frac{1}{\kappa}\left(c^*_Ac_A + \psi^*_A\psi_A\right)
\end{eqnarray}
as (up to total derivative terms)
\begin{align}
-\frac{1}{2}(\partial_+ + iA-i\hat{A})c^*_A(\partial_- - iA^*+i\hat{A}^*)c_A  &= -\frac{1}{2}D_ic^*_iD_ic_i \cr
-\frac{1}{2}(\partial_- + iA^*-i\hat{A}^*)\psi^*_A (\partial_+-iA+i\hat{A})\psi _A&=-\frac{1}{2}D_i\psi^*_AD_i\psi_A -\frac{F_{12}-\hat{F}_{12}}{2}\psi^*_A\psi_A \cr
 &= -\frac{1}{2}D_i\psi^*_AD_i\psi_A +\frac{F_{12}-\hat{F}_{12}}{2}\psi^*_A\psi_A 
\end{align}
Finally, by renaming $c_A \to (\Phi_2 , -\Phi_1, \Phi_{2'}, -\Phi_{1'})$ and $\psi_A \to (\Psi_{a=1,2},\Psi_{a'=1',2'})$, we obtain
\begin{align}
S = \int dt dx^2 &\left(\kappa A_0F_{12} + \frac{\kappa}{2} \epsilon^{ij}\partial_tA_i A_j  -\kappa\hat{A}_0\hat{F}_{12} - \frac{\kappa}{2} \epsilon^{ij}\partial_t\hat{A}_i\hat{A}_j \right. \cr
 &+  i\Phi^*_AD_0\Phi_A - \frac{1}{2}D_i\Phi^*_AD_i\Phi_A  \cr
& + i\Psi^*_AD_0\Psi_A -\frac{1}{2}D_i\Psi^*_AD_i\Psi_A \cr
&\left. - \frac{1}{2}(F_{12}-\hat{F}_{12})(\Psi_a^*\Psi_a-\Psi^*_{a'} \Psi_{a'}) \right) 
\end{align}
We see that the action is equivalent to the non-relativistic ABJM model proposed in \cite{ABJM} in the Abelian case. 

This superfield formulation reveals a hidden $SU(4)$ symmetry of the non-relativistic ABJM theory. The relativistic ABJM theory has an $SU(4)$ symmetry, while the mass deformation breaks it down to $SU(2) \times SU(2)$. As a consequence, only the $SU(2) \times SU(2)$ symmetry has been manifest in the 
original construction of the non-relativistic ABJM theory. 
However, since we treat all $\Phi_A$ and $\Psi_A$ on the same footing, it is clear that the non-relativistic Abelian ABJM theory actually has an $SU(4)$ symmetry.

\section{Non-Abelian gauge theory}
It is straightforward to generalize the Abelian superfield formulation to the non-Abelian gauge theory. Again, the gauge multiplet consists of a vector superfield $\mathcal{V}$ and a chiral superfield $\mathcal{A}$ (and its conjugate $\mathcal{A}^\dagger$). They transform as adjoint representation of the gauge group $G$.

The gauge transformations are
\begin{align}
e^\mathcal{V} &\to e^{\Lambda^\dagger} e^\mathcal{V} e^{\Lambda} \cr
e^{-\mathcal{V}} &\to e^{-\Lambda} e^{-\mathcal{V}} e^{-\Lambda^\dagger} \cr
\mathcal{A} &\to  e^{-\Lambda} (\mathcal{A} - 2i\partial_+ \Lambda) e^{\Lambda} \cr
\mathcal{A}^\dagger &\to e^{\Lambda^\dagger} (\mathcal{A}^\dagger + 2i\partial_- \Lambda^\dagger) e^{-\Lambda^\dagger} \ ,
\end{align}
where $\Lambda$ is an adjoint-valued chiral superfield.
In the following, we work in the WZ-gauge where $ \mathcal{V} = \theta \bar{\theta}2A_0$.
The remaining transformation is the usual gauge transformation for the component fields.

The supersymmetric generalization of the non-Abelian Chern-Simons action is given by
\begin{eqnarray}
S_{\mathrm{CS}} = -\frac{\kappa}{16}\int dt dx^2 d\theta^2 \mathrm{Tr}\left((\mathcal{A} + 2i\partial_+ \mathcal{V}) e^{-\mathcal{V}} (\mathcal{A}^\dagger -2i\partial_- \mathcal{V})e^{\mathcal{V}} - 2 (\partial_+ e^\mathcal{V})(\partial_- e^{-\mathcal{V}}) \right) \ .
\end{eqnarray}
The Lagrangian density is supersymmetric gauge invariant up to total derivative terms.
In the component form, the action is given by
\begin{eqnarray}
S_{\mathrm{CS}} = \frac{\kappa}{2}\int dt dx^2 \left(\epsilon^{\mu\nu\rho}\mathrm{Tr}(A_\mu \partial_\nu A_{\rho}) + \frac{2i}{3}\epsilon^{\mu\nu\rho}\mathrm{Tr}(A_{\mu}A_{\nu}A_{\rho})-\frac{1}{8} \mathrm{Tr}\lambda\lambda^\dagger \right) \ .
\end{eqnarray}

The matter multiplet (say, the fundamental representation) can be introduced by a chiral superfield $\Phi$ transforming as $\Phi \to e^{-\Lambda} \Phi$ and associated fermionic chiral superfield $\Psi$ transforming as $\Psi \to \Psi e^{\Lambda}$. We further impose the gauge covariant constraint
\begin{eqnarray}
D(e^{\mathcal{V}} \Phi) = \partial_-{\Psi}^\dagger + \frac{i}{2} \mathcal{A}^\dagger {\Psi}^\dagger \ . \label{nAc}
\end{eqnarray}
The simple matter action 
\begin{eqnarray}
 S_{\mathrm{matter}} = -\frac{1}{2}\int dt d^2x d^2\theta \left(\Phi^\dagger e^{\mathcal{V}}\Phi - {\Psi} e^{-\mathcal{V}} {\Psi}^\dagger \right)  
\end{eqnarray}
with the constraint \eqref{nAc}
gives a non-Abelian generalization of the model discussed in section 2.

\subsection{Non-Abelian non-relativistic ABJM}
As a final application, we would like to derive the non-Abelian non-relativistic ABJM theory in the superfield formulation so far developed in this paper. It has a manifest dynamical supersymmetry and we confirm the non-relativistic limit taken in \cite{ABJM} in a manifestly supersymmetric way.

We first introduce $U(N) \times U(N)$ gauge connection with the gauge transformation
\begin{align}
e^\mathcal{V} &\to e^{\Lambda^\dagger} e^\mathcal{V} e^{\Lambda} \cr
e^{\hat{\mathcal{V}}} &\to e^{\hat{\Lambda}^\dagger} e^{\hat{\mathcal{V}}} e^{\hat{\Lambda}} \cr
\mathcal{A} &\to e^{-\Lambda} (\mathcal{A}- 2i\partial_+ \Lambda) e^{\Lambda} \cr\hat{\mathcal{A}} &\to e^{-\hat{\Lambda}} (\hat{\mathcal{A}} - 2i\partial_+ \hat{\Lambda}) e^{\hat{\Lambda}}  \ .
\end{align}
The matter chiral  multiplets $(\Phi_a, \Psi_a)$ and $(\Phi_{a'}, \Psi_{a'})$ transform as
\begin{align}
\Phi_a &\to e^{-\hat{\Lambda}} \Phi_a e^{{\Lambda}} \cr
\Psi_a &\to e^{-{\Lambda}} \Psi_a e^{\hat{\Lambda}} \cr 
\Phi_{a'} & \to e^{-{\Lambda}} \Phi_{a'} e^{\hat{\Lambda}} \cr
\Psi_{a'}  & \to e^{-\hat{\Lambda}} \Psi_{a'} e^{{\Lambda}} \ ,
\end{align}
where we have treated two indices $(a=1,2)$ and $(a'=1',2')$ differently, so there remains only $SU(2)\times SU(2)$ symmetry unlike the Abelian case in section 3. 

The Chern-Simons part of the action is given by
\begin{align}
S_{\mathrm{CS}} = \int dt dx^2 d\theta^2 &-\frac{\kappa}{16}\mathrm{Tr}\left((\mathcal{A} + 2i\partial_+ \mathcal{V}) e^{-\mathcal{V}} (\mathcal{A}^\dagger -2i\partial_- \mathcal{V})e^{\mathcal{V}} - 2 (\partial_+ e^\mathcal{V})(\partial_- e^{-\mathcal{V}}) \right) \cr
&+\frac{\kappa}{16}\mathrm{Tr}\left((\hat{\mathcal{A}} + 2i\partial_+ \hat{\mathcal{V}}) e^{-\hat{\mathcal{V}}} (\hat{\mathcal{A}}^\dagger -2i\partial_- \hat{\mathcal{V}})e^{\hat{\mathcal{V}}} - 2 (\partial_+ e^{\hat{\mathcal{V}}})(\partial_- e^{-\hat{\mathcal{V}}}) \right) \cr
 = \int dt dx^2 &\frac{\kappa}{2}\left(\epsilon^{\mu\nu\rho}\mathrm{Tr}(A_\mu \partial_{\nu}A_{\rho}) + \frac{2i}{3}\epsilon^{\mu\nu\rho}\mathrm{Tr}(A_{\mu}A_{\nu}A_{\rho})-\frac{1}{8} \mathrm{Tr}\lambda\lambda^\dagger \right) \cr
&- \frac{\kappa}{2} \left(\epsilon^{\mu\nu\rho}\mathrm{Tr}(\hat{A}_\mu \partial_\nu\hat{A}_{\rho}) + \frac{2i}{3}\epsilon^{\mu\nu\rho}\mathrm{Tr}(\hat{A}_{\mu}\hat{A}_{\nu}\hat{A}_{\rho})-\frac{1}{8} \mathrm{Tr}\hat{\lambda}\hat{\lambda}^\dagger \right)  \ .
\end{align}

The matter action is given by the simple  form
\begin{align}
S_{\mathrm{matter}} = -\frac{1}{2} \int dt dx^2 d\theta^2 \mathrm{Tr}\left(\Phi_a^\dagger  e^{\hat{\mathcal{V}}}\Phi_a e^{-{\mathcal{V}}} + \Psi_a^\dagger e^{{\mathcal{V}}} \Psi_a e^{-\hat{\mathcal{V}}} + \Phi_{a'}^\dagger e^{\mathcal{V}} \Phi_{a'} e^{-\hat{\mathcal{V}}} - \Psi_{a'}^\dagger e^{\hat{\mathcal{V}}} \Psi_{a'} e^{-{\mathcal{V}}} \right) 
\end{align}
with the following constraints:
\begin{align}
D(e^{\hat{\mathcal{V}}}\Phi_{a} e^{-{\mathcal{V}}}) & = \partial_- \Psi^\dagger_{a} + \frac{i}{2} \hat{\mathcal{A}}^\dagger \Psi^\dagger_a -\frac{i}{2} {\Psi}_a^\dagger{\mathcal{A}}^\dagger \cr
D(e^{\hat{\mathcal{V}}}\Psi_{a'} e^{-{\mathcal{V}}}) &= \partial_-\Phi^\dagger_{a'} + \frac{i}{2}\hat{\mathcal{A}}^\dagger \Phi^\dagger_{a'} - \frac{i}{2}\Phi_{a'}^\dagger {\mathcal{A}}^\dagger \ . 
\end{align}
In components, we have
\begin{align}
\chi_a &= \partial_- \psi^\dagger_a + i \hat{A}^\dagger \psi_a^\dagger - i \psi_a^\dagger {A}^\dagger \cr
-2i\partial_t \phi_a + 2\hat{A}_0 \phi_a -2\phi_a {A}_0 &= \partial_- \rho^\dagger_a + i\hat{A}^\dagger \rho^\dagger_a - i\rho_a^\dagger {A}^\dagger -\frac{i}{2} \hat{\lambda}^\dagger \psi^\dagger_a + \frac{i}{2}\psi^\dagger_a {\lambda}^\dagger \cr
\rho_{a'} &= \partial_- \phi^\dagger_{a'} + i \hat{A}^\dagger \phi_{a'}^\dagger - i \phi_{a'}^\dagger {A}^\dagger \cr
-2i\partial_t \psi_{a'} + 2\hat{A}_0 \psi_{a'} -2\psi_{a'} {A}_0 &= -\partial_- \chi^\dagger_{a'} - i\hat{A}^\dagger \chi^\dagger_{a'} + i\chi_{a'}^\dagger {A}^\dagger -\frac{i}{2} \hat{\lambda}^\dagger \phi^\dagger_{a'} + \frac{i}{2}\phi^\dagger_{a'} {\lambda}^\dagger
\end{align}

The matter Lagrangian with the constraints imposed by the Lagrange multipliers is written as
\begin{align}
\mathrm{Tr}&\left( i\phi^\dagger_a D_t \phi_a-\frac{1}{2}D_-\psi_a^\dagger D_+ \psi_a + i\psi_a^\dagger D_t \psi_a + \frac{1}{2} \rho^\dagger_a \rho_a \right. \cr
+& c_a( iD_t \phi_a + \frac{1}{2} D_- \rho_a^\dagger - \frac{i}{4} \hat{\lambda}^\dagger \psi_a^\dagger + \frac{i}{4} \psi^\dagger_a {\lambda}^\dagger ) +(-iD_t \phi^\dagger_a + \frac{1}{2} D_+ \rho_a + \frac{i}{4}\psi_a\hat{\lambda} -\frac{i}{4}{\lambda} \psi_a) c^\dagger_a \cr
+&i\psi^\dagger_{a'}D_t \psi_{a'} + \frac{1}{2}\chi^\dagger_{a'} \chi_{a'}  - i\psi_{a'}^\dagger D_t \psi_{a'} -\frac{1}{2}D_-\phi^\dagger_{a'} D_+\phi_{a'} \cr
+&\left. \xi_{a'} (iD_t\psi_{a'} -\frac{1}{2}D_-\chi_{a'}^\dagger -\frac{i}{4}\hat{\lambda}^\dagger \phi_{a'}^\dagger + \frac{i}{4}\phi_{a'}^\dagger {\lambda}^\dagger) + (-iD_t\psi^\dagger_{a'} -\frac{1}{2}D_+\chi_{a'} + \frac{i}{4} \phi_{a'}\hat{\lambda} - \frac{i}{4}{\lambda} \phi_{a'} ) \xi^\dagger_{a'} \right) \cr
\end{align}
where $c_a$ are bosonic Lagrange multipliers and  $\xi_{a'}$ are fermionic ones. 
We can eliminate the auxiliary fields $\lambda, \hat{\lambda}, \rho_a, \chi_{a'}$ as well as $\phi_a, \psi_{a'}$  as
\begin{align}
\frac{\kappa}{16}\hat{\lambda} + \frac{i}{4}\psi_a^\dagger c_a -\frac{i}{4}\phi^\dagger_{a'} \xi_{a'} &= 0 \cr
-\frac{\kappa}{16} {\lambda} + \frac{i}{4} c_a\psi^\dagger_a + \frac{i}{4} \xi_{a'} \phi_{a'}^\dagger &= 0 \cr
\rho_{a} - D_- c_a &= 0 \cr
\chi_{a'} - D_- \xi_{a'} &= 0 \cr
iD_t \phi_a + iD_t c^\dagger_a &= 0 \cr
-iD_t \psi_{a'} + iD_t \xi^\dagger_{a'} &= 0 \ .
\end{align}
Substituting them back into the matter Lagrangian, we obtain the matter Lagrangian:
\begin{align}
\mathrm{Tr}&\left( i c_a^\dagger D_t c_a -\frac{1}{2}D_+c_a^\dagger D_- c_a + i \psi^\dagger_a D_t \psi_a -\frac{1}{2}D_-\psi_a^\dagger D_+ \psi_a \right. \cr 
&+i\phi_{a'}^\dagger D_t \phi_{a'} -\frac{1}{2}D_-\phi^\dagger_{a'} D_+\phi_{a'} + i\xi^\dagger_{a'} D_t \xi_{a'} -\frac{1}{2}D_+\xi^\dagger_{a'} D_-\xi_{a'}  \cr
&- \left. \frac{1}{\kappa} (\psi_a^\dagger c_a -\phi^\dagger_{a'} \xi_{a'})(c_b^\dagger\psi_b - \xi^\dagger_{b'} \phi_{b'}) +\frac{1}{\kappa}(c_a\psi^\dagger_a-\xi_{a'} \phi^\dagger_{a'})(\psi_ac_a^\dagger -\phi_{a'} \xi^\dagger_{a'}) \right) 
\end{align}

In order to compare it with the one presented in \cite{ABJM}, we use the Gauss-law constraint\begin{align}
{F}_{12} & = \frac{1}{\kappa}\left(c_a c^\dagger_a -\psi_a\psi^\dagger_a + \phi_{a'}\phi^\dagger_{a'} - \xi_{a'} \xi^\dagger_{a'} \right) \cr
\hat{F}_{12}& =\frac{1}{\kappa}\left(c^\dagger_a c_a + \psi^\dagger_a \psi_a + \phi^\dagger_{a'} \phi_{a'} +\xi^\dagger_{a'} \xi_{a'} \right)  
\end{align}
 to rewrite the kinetic term for scalars as
\begin{align}
-\frac{1}{2} \mathrm{Tr}(D_+ c_a^\dagger D_- c_a) =& -\frac{1}{2} \mathrm{Tr}(D_i c_a^\dagger D_i c_a) - \frac{1}{2} \mathrm{Tr}(c^\dagger_a c_a \hat{{F}}_{12} ) + \frac{1}{2}\mathrm{Tr}(c_a^\dagger {F}_{12} c_a)\cr
 =&-\frac{1}{2} \mathrm{Tr}(D_i c_a^\dagger D_i c_a) -\frac{1}{2\kappa}\mathrm{Tr}\left(c^\dagger_ac_a(c^\dagger_b c_b + \psi^\dagger_b \psi_b + \phi^\dagger_{b'} \phi_{b'} +\xi^\dagger_{b'} \xi_{b'}) \right) \cr
 &+ \frac{1}{2\kappa} \mathrm{Tr}\left(c_a^\dagger(c_b c^\dagger_b-\psi_b\psi^\dagger_b + \phi_{b'}\phi^\dagger_{b'} - \xi_{b'} \xi^\dagger_{b'})c_a \right) 
 \cr
-\frac{1}{2} \mathrm{Tr}(D_- \phi_{a'}^\dagger D_+ \phi_{a'}) =& -\frac{1}{2} \mathrm{Tr}(D_i \phi_{a'}^\dagger D_i \phi_{a'}) + \frac{1}{2} \mathrm{Tr}(\phi^\dagger_{a'} \phi_{a'} \hat{{F}}_{12} ) - \frac{1}{2}\mathrm{Tr}(\phi_{a'}^\dagger {F}_{12} \phi_{a'}) \cr
 =&-\frac{1}{2} D_i \phi_{a'}^\dagger D_i \phi_{a'}^\dagger +\frac{1}{2\kappa}\mathrm{Tr}\left(\phi^\dagger_{a'}\phi_{a'}(c^\dagger_b c_b + \psi^\dagger_b \psi_b + \phi^\dagger_{b'} \phi_{b'} +\xi^\dagger_{b'} \xi_{b'}) \right) \cr
 &- \frac{1}{2\kappa} \mathrm{Tr}\left(\phi_{a'}^\dagger(c_b c^\dagger_b-\psi_b\psi^\dagger_b + \phi_{b'}\phi^\dagger_{b'} - \xi_{b'} \xi^\dagger_{b'})\phi_{a'} \right) 
\end{align}
Note that the scalar potential solely comes from this rewriting and it is summarized as
\begin{align}
V_{\mathrm{bos}} =-\frac{1}{2\kappa} \mathrm{Tr}\left(\phi_{a}\phi_{[a}^\dagger \phi_{b} \phi^\dagger_{b]} - \phi_{a'} \phi^\dagger_{[a'}\phi_{b'} \phi^\dagger_{b']} \right) \ ,
\end{align}
where we have renamed $c_a \to i\phi_{a}$ in order to adjust to the convention used in \cite{ABJM}.

Finally, we rename the fermionic fields: $\psi_a \to i\epsilon_{ab} \psi_b$, $\xi_{a'} \to i\epsilon_{a'b'} \psi_{b'}$.
 With this renaming, the action completely agrees with that in \cite{ABJM} (with the replacement $k = 2\pi \kappa$ and $m = 1$):
\begin{align}
S_{\mathrm{ABJM}} = \int dt d^2x \left(L_{\mathrm{CS}} + L_{\mathrm{kin}} - V_{\mathrm{bos}} - V_{\mathrm{fer}}\right) \ ,
\end{align}
where
\begin{align}
L_{\mathrm{CS}} =&\frac{\kappa}{2}\left(\epsilon^{\mu\nu\rho}\mathrm{Tr}(A_\mu \partial_\nu A_\rho) + \frac{2i}{3}\epsilon^{\mu\nu\rho}\mathrm{Tr}(A_{\mu}A_{\nu}A_{\rho})- \epsilon^{\mu\nu\rho}\mathrm{Tr}(\hat{A}_\mu \partial_\nu\hat{A}_{\rho}) - \frac{2i}{3}\epsilon^{\mu\nu\rho}\mathrm{Tr}(\hat{A}_{\mu}\hat{A}_{\nu}\hat{A}_{\rho}) \right)  \cr
L_{\mathrm{kin}} =& \mathrm{Tr}\left( i \phi_a^\dagger D_t \phi_a -\frac{1}{2}D_i\phi_a^\dagger D_i \phi_a + i \psi^\dagger_a D_t \psi_a -\frac{1}{2}D_-\psi_a^\dagger D_+ \psi_a \right. \cr 
& \left. +i\phi_{a'}^\dagger D_t \phi_{a'} -\frac{1}{2}D_i\phi^\dagger_{a'} D_i\phi_{a'} + i\psi^\dagger_{a'} D_t \psi_{a'} -\frac{1}{2}D_+\psi^\dagger_{a'} D_-\psi_{a'}\right) \cr
V_{\mathrm{bos}} =& -\frac{1}{2\kappa} \mathrm{Tr}\left(\phi_{a}\phi_{[a}^\dagger \phi_{b} \phi^\dagger_{b]} - \phi_{a'} \phi^\dagger_{[a'}\phi_{b'} \phi^\dagger_{b']} \right) \cr
V_{\mathrm{fer}} =&-\frac{1}{2\kappa} \mathrm{Tr} \left[ (\phi^\dagger_a\phi_a + \phi^\dagger_{a'} \phi_{a'})(\psi^{\dagger}_{b}\psi_b -  \psi^{\dagger}_{b'} \psi_{b'}) +(\phi_a\phi^\dagger_a +\phi_{a'}\phi^\dagger_{a'} )(\psi_b \psi^{\dagger}_{b} -  \psi_{b'} \psi^{\dagger}_{ b'} ) \right. \cr
&- 2 \phi_a\phi^\dagger_b \psi_a\psi^{\dagger}_{b} + 2 \phi_{a'} \phi^\dagger_{b'} \psi_{a'} \psi^{\dagger}_{b'} - 2 \phi^\dagger_a \phi_b\psi^{\dagger}_{a}\psi_b +2 \phi^\dagger_{a'} \phi_{b'} \psi^{\dagger}_{a'} \psi_{b'} \cr
&  -i\epsilon^{ab}\epsilon^{c'd'} \phi^\dagger_a \psi_b \phi^\dagger_{c'} \psi_{d'} - i \epsilon^{bc}\epsilon^{a'd'}\phi^\dagger_{a'} \psi_b \phi^\dagger_c\psi_{d'}+i\epsilon^{a'b'}\epsilon^{cd} \phi^\dagger_{a'} \psi_{b'} \phi^\dagger_{c} \psi_{d}+ i\epsilon^{b'c'}\epsilon^{ad}\phi^\dagger_{a} \psi_{b'}\phi^\dagger_{c'}\psi_{d} \cr
& \left. +i\epsilon^{ab}\epsilon^{c'd'} \phi_a \psi^{\dagger}_{b} \phi_{c'} \psi^{\dagger}_{d'} + i \epsilon^{bc}\epsilon^{a'd'}\phi_{a'} \psi^{\dagger}_{b} \phi_c\psi^{\dagger}_{d'}-i\epsilon^{a'b'}\epsilon^{cd} \phi_{a'} \psi^{\dagger}_{b'} \phi_{c} \psi^{\dagger}_{d}- i\epsilon^{b'c'}\epsilon^{ad}\phi_{a} \psi^{\dagger}_{b'}\phi_{c'}\psi^{\dagger}_{d}    
\right] 
\end{align}

\section{Discussions}
In this paper, we have developed a superfield formulation for non-relativistic Chern-Simons-Matter theories in $(1+2)$ dimension. We have successfully reproduced the non-relativistic Chern-Simons-Matter theories proposed in the literatures while manifestly preserving the dynamical supersymmetry.

Our formulation is minimal in the sense that it only possesses manifest dynamical supersymmetry, which has both advantage and disadvantage. The disadvantage is that we fail in manifesting some additional symmetries such as Poincar\'e invariance or Galilean invariance. On the other hand, this minimal structure allows us to study the theory with no such additional structures. For instance, we can easily introduce the supersymmetric extensions of the Maxwell term by adding $S_E + S_M$ with
\begin{align}
S_{\mathrm{E}} &= \frac{1}{16 g_E^2}\int dt d^2x d^2\theta \left(D(\mathcal{A} + 2i\partial_+\mathcal{V})\bar{D}(\mathcal{A}^*-2i\partial_- \mathcal{V}) \right)\cr
S_{\mathrm{M}}& = -\frac{1}{2g_M^2} \int dt d^2x d^2 \theta \Gamma^* \Gamma \ ,
\end{align}
where we impose the constraint for a fermionic chiral multiplet $\Gamma$ as
\begin{eqnarray}
D \Gamma + \bar{D} \Gamma^* = -i(\partial_- \mathcal{A} -\partial_+\mathcal{A}^* + 2i \partial_-\partial_+\mathcal{V}) \ . \label{constmx}
\end{eqnarray}
Similarly, we could study the supersymmetric extension of the Ho\v{r}ava-Lifshitz term, where we need two additional fermionic chiral multiplets $\Gamma_i$ with the constraint\footnote{As a consequence, the ``gaugino" $\zeta_i = \partial_\theta \Gamma_i|_{\theta =0}$ has a vector indices $i = 1,2$. This has been independently observed by C.~M.~Thompson.}
\begin{eqnarray}
D \Gamma_i + \bar{D} \Gamma^*_i = -i\partial_i(\partial_- \mathcal{A} -\partial_+\mathcal{A}^* + 2i \partial_-\partial_+\mathcal{V}) \ . \label{consthl}
\end{eqnarray}
The correspongin action is
\begin{eqnarray}
S_{\mathrm{HL}} = -\alpha \int dt d^2x d^2 \theta \Gamma_i^* \Gamma_i \ .
\end{eqnarray}

In this paper,  we have shown the manifestly supersymmetric form of the non-relativistic ABJM theory with 14 supercharges.
We can study different (less supersymmetric) non-relativistic limits of the ABJM model presented in \cite{ABJM}. In order to realize different limits in our superfield approach, we simply change the constraint for $(a')$ multiplets: $D(e^\mathcal{V} \Phi_{a'} e^{-\hat{\mathcal{V}}}) = \partial_- \Psi^\dagger_a + \frac{i}{2}\mathcal{A}^\dagger \Psi^\dagger_{a'} - \frac{i}{2} \Psi^\dagger_{a'} \hat{\mathcal{A}}^\dagger$. The matter content is exactly the same as that studied in section 4 of \cite{ABJM}:
\begin{align}
L_{\mathrm{kin}} =& \mathrm{Tr}\left( i \phi_a^\dagger D_t \phi_a -\frac{1}{2}D_+\phi_a^\dagger D_- \phi_a + i \psi^\dagger_a D_t \psi_a -\frac{1}{2}D_-\psi_a^\dagger D_+ \psi_a \right. \cr 
& \left. +i\hat{\phi}_{a'}^\dagger D_t \hat{\phi}_{a'} -\frac{1}{2}D_+\hat{\phi}^\dagger_{a'} D_-\hat{\phi}_{a'} + i\hat{\psi}^\dagger_{a'} D_t \hat{\psi}_{a'} -\frac{1}{2}D_-\hat{\psi}^\dagger_{a'} D_+\hat{\psi}_{a'}\right)
\end{align}
 However, we note that the potential is different:\footnote{For example, the terms like $\epsilon^{ab} \epsilon^{c'd'}\phi^\dagger_a \psi_b \hat{\phi}^\dagger_{c'} \hat{\psi}_{d'}$ are missing in the action presented in section 4.2 of \cite{ABJM} even if we rewrite the potential by using the Gauss-law constraint. This was independently pointed out by M.~Sakaguchi.}
\begin{align}
V = &\frac{1}{\kappa}(\psi_a^\dagger \phi_a + \hat{\phi}_{a'} \hat{\psi}^\dagger_{a'})(\phi^\dagger_a\psi_a + \hat{\psi}_{a'} \hat{\phi}^\dagger_{a'}) - \frac{1}{\kappa}(\phi_a\psi_a^\dagger + \hat{\psi}^\dagger_{a'} \hat{\phi}_{a'})(\psi_a\phi^\dagger_a +  \hat{\phi}^\dagger_{a'} \hat{\psi}_{a'}) 
\end{align} 
Actually, there is another possibility: we exchange the representation of $\Phi_{a'}$ with that of $\Psi_{a'}$ and introduce the constraint $D(e^{\mathcal{V}}\Psi_{a'}e^{-\hat{\mathcal{V}}}) = \partial_- \Phi^\dagger_{a'} + \frac{i}{2} \mathcal{A}^\dagger \Phi^\dagger_{a'} - \frac{i}{2} \Phi^\dagger_{a'} \hat{\mathcal{A}}^\dagger$. This gives a different result, but it again shows potential terms that do not arise in the non-relativistic limit of ABJM theory.

 All these are consistent because it was shown that the non-relativistic limit taken in section 4 of \cite{ABJM} only preserves the kinematical supersymmetry. The superfield formulation here shows that  there exists a deformation of the potential so that the dynamical supersymmetry is preserved. Incidentally, it is this deformed non-relativistic ABJM theory whose index was computed in \cite{Nakayama:2008qm}. It would be interesting to give clear physical understanding of this deformation from the viewpoint of the original ABJM theory.

\section*{Acknowledgements}
The author would like to thank M.~Sakaguchi, C.~M.~Thompson and K.~Yoshida for fruitful discussions. The work was supported in part by the National Science Foundation under Grant No.\ PHY05-55662 and the UC Berkeley Center for Theoretical Physics.

\end{document}